\documentclass[aps,preprint,showkeys,showpacs,superscriptaddress,onecolumn,floatfix]{revtex4}
\usepackage{amsfonts}
\usepackage{amsmath}
\usepackage{amssymb}
\usepackage{graphicx}
\usepackage{subfigure}
\usepackage{color}
\def\be{\begin{equation}}
\def\ee{\end{equation}}
\def\bea{\begin{eqnarray}}
\def\eea{\end{eqnarray}}
\def\bg{\begin{eqnarray}}
\def\en{\end{eqnarray}}

\def\ra{\rightarrow}

\begin{document}
\preprint{JLAB-THY-08-780}
\title{Interplay of Spin and Orbital Angular Momentum in the Proton  
\footnotetext{Notice: Authored by Jefferson Science Associates, LLC under U.S. DOE Contract No. DE-AC05-06OR23177. The U.S. Government retains a non-exclusive, paid-up, irrevocable, world-wide license to publish or reproduce this manuscript for U.S. Government purposes.}}
\author{Anthony W. Thomas}
\affiliation{Thomas Jefferson National Accelerator Facility, 12000 Jefferson Ave., Newport News, VA 23606, USA}
\affiliation{College of William and Mary, Williamsburg VA 23187, USA}
\affiliation{CSSM, School of Chemistry and Physics, University of Adelaide, Adelaide SA 5005 Australia}
\begin{abstract}
We derive the consequences of the Myhrer-Thomas explanation of the 
proton spin problem for the distribution of orbital angular 
momentum on the valence and sea quarks. After QCD evolution these 
results are found to be in very good agreement with both recent 
lattice QCD calculations and the experimental constraints from 
Hermes and JLab.
\end{abstract}
\date{\today}
\maketitle
\section{Introduction}
There is no more fundamental question concerning the 
structure of the nucleon than the distribution of 
spin and orbital angular momentum over its 
quarks and gluons~\cite{Thomas:2001kw,Bass:2007zz}. 
This issue has been of 
enormous topical interest since the European Muon 
Collaboration (EMC) reported that most of the nucleon 
spin was not carried as the spin of its quarks and 
anti-quarks~\cite{Ashman:1987hv}. 
Over the last 20 years there has been 
tremendous progress in unravelling this mystery. In 
particular, it is now known that the missing spin 
fraction is of order 
2/3~\cite{Airapetian:2007mh,Alexakhin:2006vx}, 
rather than 90\% and 
furthermore the contribution from polarized gluons 
is less than 5\% (corresponding to $| \Delta G| 
< 0.3$~\cite{PHENIX_1,Adare:2007dg,STAR_1,Abelev:2007vt,Dharmawardane:2006zd,Leader:2006xc}). 
It was recently shown by Myhrer and Thomas~\cite{Myhrer:2007cf} 
that the modern spin discrepancy can be rather well 
explained in terms of standard features of the 
non-perturbative structure of the nucleon, namely 
relativistic motion of the 
valence quarks~\cite{Chodos:1974pn}, the pion 
cloud required by chiral symmetry~\cite{Schreiber:1988uw} 
and an exchange 
current contribution associated with the 
one-gluon-exchange hyperfine interaction~\cite{Myhrer:1988ap}.

Here we derive the consequences of the Myhrer-Thomas 
work for the distribution of orbital angular momentum 
on the quarks and anti-quarks. These results are then 
tested against the latest measurements of the 
Generalized Parton Distributions from Hermes and 
JLab, as well as lattice QCD. 
We shall see that once the appropriate connection between the quark model 
and QCD is made at an appropriately low scale, there 
is a remarkable degree of consistency between 
all three determinations. This not only gives us 
considerable confidence in the physical picture 
provided by Myhrer and Thomas but it also provides 
much needed insight into the physical content of the 
lattice QCD simulations.

The structure of the paper is that we first track 
where, in the Myhrer-Thomas picture, the missing spin 
resides as orbital angular momentum on valence quarks 
and anti-quarks. We then recall that orbital angular 
momentum is not a renormalization group invariant
and argue, following 30 years of similar 
arguments~\cite{Parisi:1976fz,Schreiber:1991tc}, 
that the model values should be associated 
with a very low scale. Solving the QCD evolution 
equations for the up and down quark angular momenta 
then leads to the remarkable result that the orbital 
angular momentum of the up and down quarks cross over 
around 1 GeV$^2$, so that at the scale of 
current experiments or lattice QCD simulations 
$L^d$ (the orbital angular momentum carried by 
down {\em and} anti-down quarks) is positive and 
greater than $L^u$, which tends to be negative.

Consider first the relativistic motion of the valence 
quarks, described (e.g.) by solving the 
Dirac equation for a spin up particle in an s-state. 
The lower component of the corresponding spinor has 
the quark spin predominantly down (i.e. spin down 
to spin up in the ratio 
2/3:1/3), because 
the corresponding, p-wave orbital angular momentum 
is up. Thus the relativistic correction which lowers 
the quark spin fraction to about 65\%, leads to 
35\% of the proton spin being carried as valence quark 
orbital angular momentum. If, for simplicity, we 
start with an SU(6) wavefunction, the 
$u-d$ components are in the ratio +4/3:-1/3. This 
is summarized in line 2 of Table 1.

As originally derived by Hogaasen and 
Myhrer~\cite{Hogaasen:1988jd}, 
the exchange current correction to spin 
dependent quantities, such as baryon axial charges 
and magnetic moments, arising from the widely used  
one-gluon-exchange hyperfine interaction, is 
dominated by those diagrams involving excitation 
of a p-wave anti-quark. The total correction to the 
spin is $\Sigma \ra \Sigma - 3G$, where 
$G = 0.05$ involves exactly the same matrix elements. 
(N.B. We follow Hogaasen and Myhrer in using $G$ to 
denote the product of $\alpha_s$ times the relevant bag 
model matrix elements. It bears no relation to the gluonic 
parton distribution or $\Delta G$, which is traditionally 
used to denote the spin carried by polarized glue.)
In this case, the 15\% of the proton spin lost to quarks 
through this mechanism is converted to 
orbital angular momentum of the p-wave anti-quark.  
This is summarized in line 3 of Table 1.

The pion cloud of the nucleon required by chiral 
symmetry~\cite{Thomas:2007bc,Theberge:1980ye,Thomas:1982kv} 
leads to a 
multiplicative correction to 
the nucleon spin, $Z - \frac{1}{3} P_{{\rm N}\pi} 
+ \frac{5}{3} P_{\Delta \pi}$~\protect\footnote{Here $Z$ 
is the wavefunction renormalization constant, the 
probability of finding the bare quark structure of 
the nucleon, and $P_{{\rm N}\pi}$ and 
$P_{\Delta \pi}$ are, respectively, the probabilities 
of finding a pion with the 
accompanying baryon having N or $\Delta$ quantum 
numbers.} of order 0.75 to 0.80. For the 
N $\pi$ Fock component of the nucleon wavefunction 
the angular 
momentum algebra is identical to that of the lower 
component of the quark spinor mentioned above. 
That is, the pion tends to have positive (p-wave) 
orbital angular momentum, while the N spin is 
down. From the point of view of a deep inelastic probe the 
pion is (predominantly) a quark-anti-quark pair but 
since they are coupled to spin zero they contribute 
nothing to the spin structure function. 

The flavor structure of the 
pion-baryon Fock components needs 
a little care, for example the dominant N $\pi$ 
component is n $\pi^+$, so the pion orbital angular 
momentum in this case is shared by a $u$-quark 
and a $\bar{d}$-anti-quark -- leading naturally 
to an excess of $\bar{d}$ quarks in the proton 
sea~\cite{Thomas:1983fh}. The final distribution 
of spin and orbital angular momentum, obtained after 
applying the pionic correction to the relativistic 
quark model, including the effect of the 
one-gluon-exchange hyperfine interaction, is shown 
in the final line of Table 1.
\begin{table}[htbp]
\begin{center}
\caption{Distribution of the fraction 
of the spin of the nucleon 
as spin and orbital angular momentum of 
its constituent quarks at the model (low energy) scale. 
Successive lines down the 
table show the result of adding a new effect to 
all the preceding effects. (Note that for all terms 
the contributions of both quarks and anti-quarks of 
a given flavor are included.)} 
\label{spe1}
\begin{tabular}[t]{cccc}
\hline 
\hline
& $L^u$ & $L^d$ & $\Sigma$ \\
\hline
\hline
Non-relativistic & 0 & 0 & 100 \\
Relativistic & 0.46 & -0.11 & 0.65 \\
OGE & 0.67 & -0.16 & 0.49 \\
Pion cloud & 0.64 & -0.03 & 0.39 \\
\hline
\hline
\end{tabular}
\end{center}
\end{table}

The very clear physical picture evident from Table 1 
is that the spin of the proton resides predominantly 
as orbital angular momentum of the $u$ 
(and $\bar{u}$) quarks. In contrast, 
the $d$ (and $\bar{d}$) quarks 
carry essentially no orbital angular momentum.
The total angular momentum is shared between 
the $u$ (and $\bar{u}$)
quarks, $J^u$, and the $d$ (and $\bar{d}$) 
quarks, $J^d$, in the ratio 
$J^u : J^d = 0.74 : -0.24$. (Note that there are 
no strange quarks in the Myhrer-Thomas calculation, 
so $\Sigma$ in Table 1 is $\Delta u + \Delta d$. 
Combining this with $g_A^3 \equiv \Delta u - \Delta d 
= 1.27$ yields these values. A more sophisticated treatment, 
including the $KN$ Fock component of the proton 
wavefunction~\cite{Signal:1987gz}, would lead 
to a very small non-zero value 
of $\Delta s$~\cite{Melnitchouk:1999mv}.)

At first appearance, these results seem to disagree 
with the first indications from lattice 
QCD~\cite{Hagler:2007xi,Richards:2007vk}, which 
suggest that $L^d$ tends to be positive, while 
$L^u$ is negative. One should observe that
these calculations were performed at fairly large quark mass 
and omit disconnected terms, which may 
carry significant orbital angular 
momentum~\cite{Mathur:1999uf} and 
are certainly needed to account for the U(1) axial 
anomaly. Nevertheless, the apparent discrepancy 
is of concern. 

At this point, we recall the crucial fact that 
neither the total, nor the orbital angular momentum 
is renormalization group invariant 
(RGI)~\cite{Ji:1995cu}. The lattice 
QCD values are evaluated at a scale set by 
the lattice spacing, around 4 GeV$^2$. On the other 
hand, we have not identified the scale
corresponding to the values derived in our 
chiral quark model. Indeed, there is no unambiguous  
way to do so unless the model can be derived 
rigorously from non-perturbative QCD. 

This problem has been considered for 
more than 30 years~\cite{Parisi:1976fz}, driven 
initially by the fact that in a typical, 
valence dominated quark model, the fraction of 
momentum carried by the valence quarks is near 
100\%, whereas at 4 GeV$^2$ the experimentally 
measured fraction is nearer 35\%. Given that 
QCD evolution implies that the momentum carried 
by valence quarks is a monotonically decreasing 
function of the scale, the only place to match 
a quark model to QCD is at a low scale, $Q_0$. Early 
studies within the bag model found this scale to 
be considerably less than 1 GeV~\cite{Schreiber:1991tc}.

Over the last decade, 
this idea has been used with remarkable 
success to describe the data from HERA, over an 
enormous range of $x$ and $Q^2$, starting from a 
valence dominated set of input parton
distributions at a scale of 
order 0.4 GeV~\cite{Gluck:1993im}. A 
similar scale is needed to match parton distributions 
calculated in various modern quark models to 
experimental data~\cite{Cloet:2007em}. 
We note that the comparison 
between theory and experiment after QCD evolution 
is not very sensitive to the order of perturbation 
theory at which one works. However, what does change is 
the unphysical starting scale. For this reason 
we present results here at leading order - which 
also avoids questions of scheme dependence.

The QCD evolution equations for angular momentum 
in the flavor singlet case were studied by 
Ji, Tang and Hoodbhoy~\cite{Ji:1995cu}. 
The scheme used corresponds to the 
choice of a renormalization scheme which preserves 
chiral symmetry, rather than gauge 
symmetry~\cite{Adler:1969gk,Crewther:1978zz}, so that 
$\Sigma$ is scale invariant. The gluon spin then
takes the form:
\be
\Delta G(t) = - \frac{4 \Sigma}{\beta_0} +
\frac{t}{t_0} \left( \Delta G(t_0) + 
\frac{4 \Sigma}{\beta_0} \right) \, ,
\ee 
where $t = \ln (Q^2/ \Lambda^2_{\rm QCD})$ and 
$\alpha_s(Q^2) = 4 \pi 
/ [\beta_0 \ln (Q^2 / \Lambda^2_{\rm QCD}]$, with
$\beta_0 = 11 - 2 N_f/3$.     
The total quark and gluon orbital angular momenta 
satisfy coupled differential equations with 
solutions which can be written in closed form:
\bea
L^{u+d+s}(t) &+& \frac{\Sigma}{2} = \frac{1}{2} 
\frac{3 N_f}{16 + 3 N_f} + 
\left( \frac{t}{t_0} \right)^{- \frac{32+6N_f}{9 \beta_0}} 
\left( L^{u+d}(t_0) + \frac{\Sigma}{2} -
\frac{3 N_f}{16 + 3 N_f} \right)  
\nonumber \\
L^g(t) &=& - \Delta G(t) + \frac{1}{2} 
\frac{16}{16 + 3 N_f} +
\left( \frac{t}{t_0} \right)^{- \frac{32+6N_f}{9 \beta_0}} 
\left(L^g(t_0) + \Delta G(t_0) - \frac{1}{2} 
\frac{16}{16 + 3 N_f} \right) \, .
\label{equ:LqLg}
\eea
The solution for the non-singlet case, $L^{u-d} 
\equiv L^u - L^d$, is much simpler. Note that we 
now specialize to the case 
of 3 active flavors ($N_f=3$):
\be
L^{u-d}(t) + \frac{\Delta u - \Delta d}{2} =
\left( \frac{t}{t_0} \right)^{-\frac{32}{9 \beta_0}}
\left( L^{u-d}(t_0) + \frac{\Delta u - \Delta d}{2}
\right) \, .
\label{eq:Lumd}
\ee
One can also solve for the non-singlet combination 
$L^{u+d} - 2 L^s$ and hence obtain explicit expressions for 
$L^u$ and $L^d$ (assuming, as in the Myhrer-Thomas work, 
that $\Delta s = L^s =0$ at the model scale):
\bea
L^{u(d)} &=& -\frac{\Delta u}{2} \left( -\frac{\Delta d}{2}\right) +
0.06  \nonumber \\
&+& \frac{1}{3} \left( \frac{t}{t_0} \right)^{-\frac{50}{81}} \left[
L^{u+d}(t_0) + \frac{\Sigma}{2} - 0.18 \right] \nonumber \\
&+& \frac{1}{6} \left( \frac{t}{t_0} \right)^{-\frac{32}{81}} \left[
L^{u+d}(t_0) \pm 3 L^{u - d}(t_0) \pm g_A^{(3)} + \frac{\Sigma}{2} 
\right] \, .
\label{eq:Luandd}
\eea

\begin{figure}
\begin{center}
\includegraphics[width=12cm,angle=0]{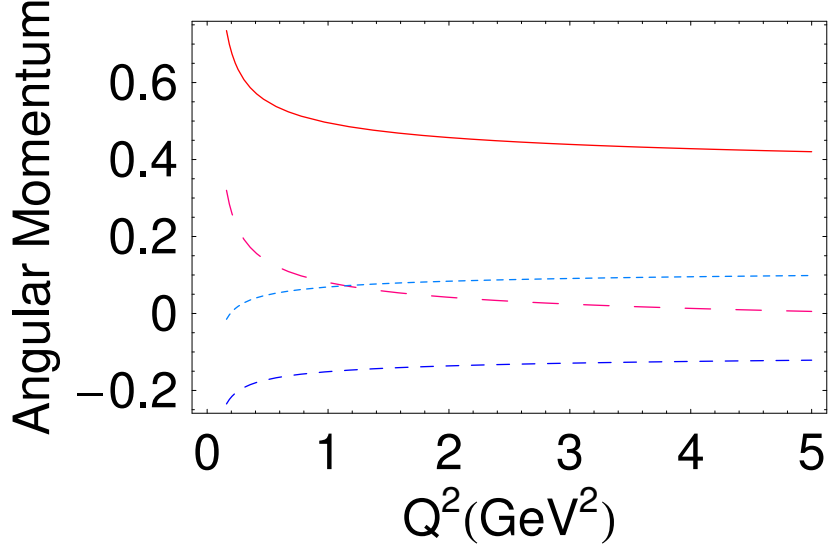}
\caption{Evolution of the total angular momentum and the orbital angular 
momentum of the up and down quarks in the proton - from top 
to bottom (at 4 GeV$^2$): 
$J^u$ (solid), $L^d$ (smallest dashes), $L^u$ (largest dashes) 
and $J^d$ (middle length dashes). 
In this case it is assumed that the 
gluons carry no spin or orbital angular momentum at the model scale (0.4 GeV).
\label{fig:1}}
\end{center}
\end{figure}

We are now in a position to evaluate the total 
and orbital angular momentum carried by each flavor 
of quark as a function of $Q^2$, given some choice 
of initial conditions. Choosing $N_f = 3$, 
$\Lambda_{\rm QCD} = 0.24$GeV and $Q_0 = 0.4$GeV, 
together with the values given in Table 1 
(and $L(t_0)=\Delta G(t_0)=0$), 
we find the results shown in Fig.~\ref{fig:1}. 
The behaviour of $J^u$ and $J^d$ is relatively 
simple, with the former decreasing fairly rapidly at 
low $Q^2$ and the latter increasing. Both settle 
down to slow variation above 1 GeV$^2$, with the 
sum around 60\% of the total nucleon spin -- the rest 
being carried as orbital angular momentum and spin 
by the gluons. A similar result has also been reported in the context
of the chiral quark soliton model~\cite{Wakamatsu:2007ar}.

While the behaviour of $J^{u,d}$ is unremarkable, 
the corresponding behaviour of $L^{u,d}$ is 
spectacular. $L^u$ is large and positive 
and $L^d$ very small and negative at the model 
scale but they very rapidly cross and settle down 
inverted above 1 GeV$^2$ ! The reason for this 
behaviour is easily understood, because asymptotically 
$L^u$ and $L^d$ tend to $0.06 - \Delta u /2$ and $0.06 - \Delta d /2$, 
or -0.36 and +0.28, respectively. This is a model 
independent result and it is simply a matter of 
how fast QCD evolution takes one from the familiar 
physics at the model scale to the asymptotic limit.
\begin{figure}
\begin{center}
\includegraphics[width=12cm,angle=0]{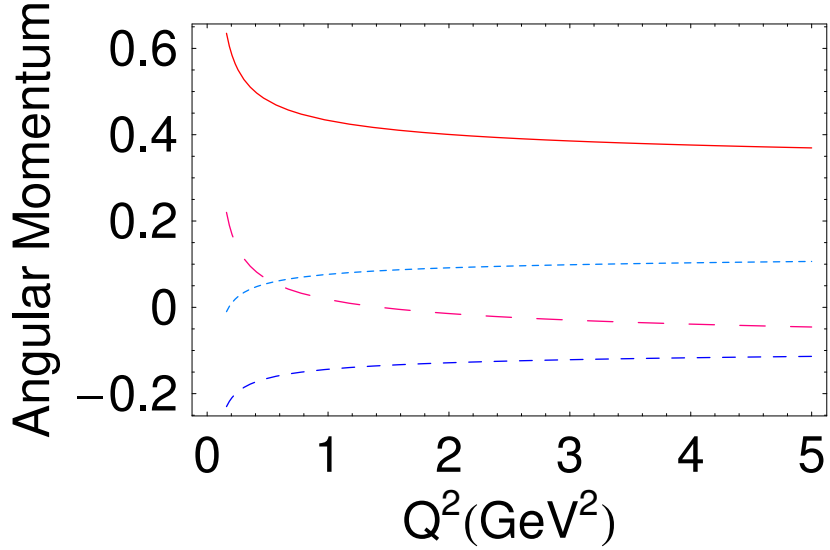}
\caption{Evolution of the total angular momentum and the orbital angular
momentum of the up and down quarks in the proton - from top
to bottom (at 4 GeV$^2$): 
$J^u$ (solid), $L^d$ (smallest dashes), $L^u$ (largest dashes)
and $J^d$ (middle length dashes). 
In this case it is assumed that the
gluons carry 0.1 units of angular momentum at the model scale (0.4 GeV).
\label{fig:2}}
\end{center}
\end{figure}

As we have already noted, the lattice QCD data for the orbital angular 
momentum carried by the $u$ and $d$ quarks has a number of systematic errors. 
Disconnected terms are as yet uncalculated and the data needs to be 
extrapolated over a large range in both pion mass and momentum transfer 
in order to extract the physical values of $J^u$ and $J^d$. 
Nevertheless, for all these cautionary remarks, the results just 
reported are consistent with the latest lattice results 
of H\"agler {\it et al.}~\cite{Hagler:2007xi}. For example, 
they report $J^{u+d}$ in the 
range 0.25 to 0.29 at the physical pion mass (their Fig.~47) in comparison 
with 0.30 in the calculation reported above. They also report $L^{u+d} 
\sim 0.06$ in comparison with 0.11 in this work. Finally, the qualitative 
feature that $L^d$ is positive and bigger than $L^u$ is, as we have explained, 
clearly reproduced in the current work.
\begin{figure}
\begin{center}
\includegraphics[width=12cm,angle=0]{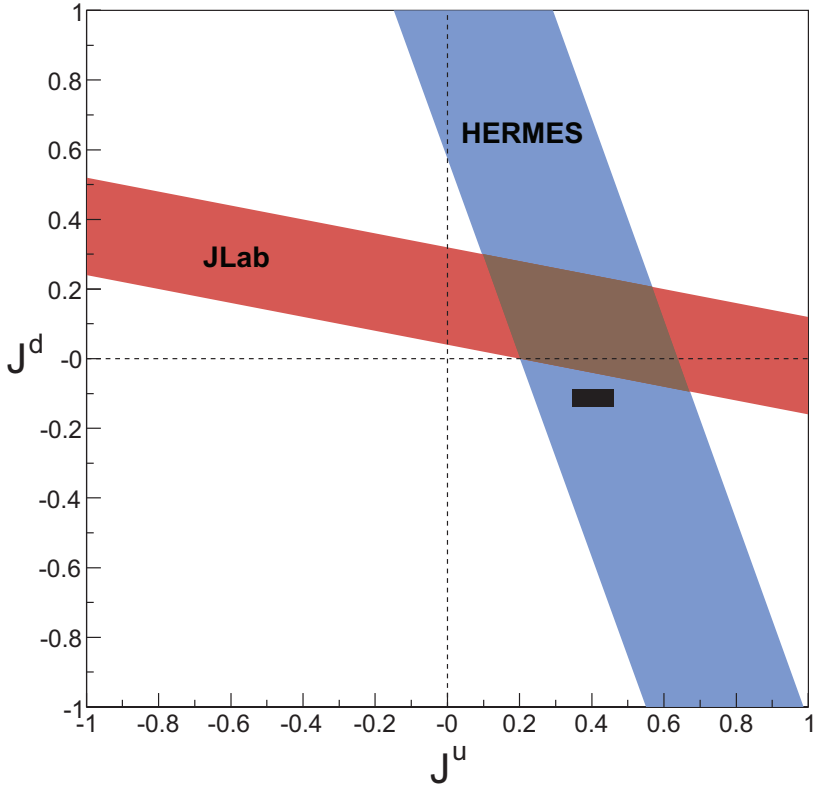}
\caption{Comparison between the constraints on the total angular momentum 
carried by $u$ and $d$ quarks in the proton, derived from experiments on 
DVCS at Hermes~\protect\cite{Ellinghaus:2005uc,Ye:2006gza} 
and JLab~\protect\cite{Mazouz:2007vj},  
and the model 
of Myhrer and Thomas (the small dark rectangle) as explained in this work.
\label{fig:DVCS}}
\end{center}
\end{figure}

Although it is clear that $\Delta G$ is too small to give a  
major correction to the spin sum rule through the axial 
anomaly~\cite{Altarelli:1988nr,Carlitz:1988ab}
(e.g. $-N_f \alpha_s \Delta G/(2 \pi) 
\sim 0.05$ for $\Delta G = 0.3$ at $Q^2 = 3$ GeV$^2$), 
it can still be non-zero and it will 
continue to be critical to pin it down more accurately. As just one example of 
the effect of a small gluon spin fraction at the model scale, 
in Fig.~\ref{fig:2} we show the evolution of the angular momentum on 
the $u$ and $d$ quarks if $\Delta G$ is set to 0.1 at the starting scale 
(and $L^{u(d)}$ lowered proportionately to preserve the proton spin). While 
the qualitative behaviour is identical there are non-trivial quantitative 
changes. In particular, $L^u$ moves down from 0.01 to -0.03 
and $J^{u+d}$ moves down to 0.26 at 4 GeV$^2$. We note that the nature 
of the QCD evolution is such that the changes in the values of $L^u$ and 
$L^d$ at 4 GeV$^2$ are considerably smaller than the changes at the model 
scale. This has the effect of reducing the uncertainty on the predictions 
of the model at the scale where they can be compared with data.

On the experimental side, the extraction of information about the quark 
angular momentum is still in its very early stage of development. One 
needs to rely on a model to analyze the experimental data, which are still 
at sufficiently low $Q^2$ that one cannot be sure that the handbag mechanism 
really dominates. Nevertheless, the combination of DVCS data on the proton 
from Hermes~\cite{Ellinghaus:2005uc,Ye:2006gza} 
and the neutron from JLab~\cite{Mazouz:2007vj} 
provides two constraints on $J^u$ and 
$J^d$, within the model of 
Goeke {\it et al.}~\cite{Goeke:2001tz,Vanderhaeghen:1999xj}, 
as shown 
in Fig.~\ref{fig:DVCS}. Also shown there  
is the prediction of the  
present work. The uncertainties shown correspond to a few percent variation 
in the relativistic correction, a 20\% reduction in the one-gluon-exchange 
correction and the uncertainty in the pion cloud correction quoted 
by Myhrer and Thomas 
(i.e. $Z -P_{N\pi}/3 +5P_{\Delta \pi}/3 \in (0.75,0.80)$). 
It also includes the variation in the scale between 
2 and 4 GeV$^2$ and the effect of $\Delta G$ being as large as 0.1 at the 
model scale. Clearly, within the present uncertainties, most notably 
the relatively low $Q^2$ of the JLab data and the unknown model 
dependence of the extraction of $J^{u(d)}$, there 
is a remarkable degree of agreement.

In summary, we have shown that the resolution of the spin crisis proposed 
by Myhrer and Thomas, which implies that the majority of the spin of the 
proton resides on $u$ and $\bar{u}$ quarks, after QCD evolution is 
consistent with current determinations from lattice QCD and experimental data 
on deeply virtual Compton scattering. The effect of QCD evolution in 
inverting the orbital angular momentum of the $u$ and $d$ quarks in the 
model was especially important. For the future, we look forward to improvements 
in both these areas, with lattice simulations at lower quark mass including 
the elusive disconnected terms and experimental data at higher $Q^2$ and $x$, 
particularly following the 12 GeV Upgrade at JLab.

\begin{acknowledgments}
It is a pleasure to acknowledge the hospitality of Derek Leinweber 
and Anthony Williams during a visit to the CSSM, during which much 
of this work was performed.
This work was supported by DOE contract DE-AC05-06OR23177, 
under which Jefferson Science Associates, LLC, operates Jefferson Lab. 
\end{acknowledgments}
%

%
\end{document}